\documentclass[aps,prx,amsfonts,floatfix,superscriptaddress,longbibliography,twocolumn,footinbib]{revtex4-2}
\usepackage[utf8]{inputenc}
\usepackage{amsmath}    
\usepackage{amssymb}
\usepackage{mathbbol}
\usepackage{graphicx}   
\usepackage{verbatim}   
\usepackage{subfigure}  
\usepackage{hyperref} 
\usepackage{scalerel}
\usepackage{cancel}
\usepackage{soul}
\usepackage[normalem]{ulem}
\usepackage[dvipsnames]{xcolor}
\usepackage{siunitx}
\usepackage{physics}
\usepackage{enumerate} 
\usepackage{empheq}

\begin{document}
\title{Dynamical detection of extended nonergodic states in many-body quantum systems}
\author{David A. Zarate-Herrada}
\affiliation{Institute of Physics, Benem\'erita Universidad Aut\'onoma de Puebla, 72570 Puebla, Pue., Mexico}
\author{Isa\'ias Vallejo-Fabila}
\affiliation{Department of Physics, University of Connecticut, Storrs, Connecticut 06269, USA}
\author{Lea F. Santos}
\affiliation{Department of Physics, University of Connecticut, Storrs, Connecticut 06269, USA}
\author{E. Jonathan Torres-Herrera}
\affiliation{Institute of Physics, Benem\'erita Universidad Aut\'onoma de Puebla, 72570 Puebla, Pue., Mexico}

\begin{abstract}
Fractal dimensions are tools for probing the structure of quantum states and identifying whether they are localized or delocalized in a given basis. These quantities are commonly extracted through finite-size scaling, which limits the analysis to relatively small system sizes. In this work, we demonstrate that the correlation fractal dimension $D_2$ can be directly obtained from the long-time dynamics of interacting many-body quantum systems. Specifically, we show that it coincides with the exponent of the power-law decay of the time-averaged survival probability, defined as the fidelity between an initial state and its time-evolved counterpart. This dynamical approach avoids the need for scaling procedures and enables access to larger systems than those typically reachable via exact diagonalization. We test the method on various random matrix ensembles, including full random matrices, the Rosenzweig-Porter model, and power-law banded random matrices, and extend the analysis to interacting many-body systems described by the one-dimensional Aubry-Andr\'e model and the disordered spin-1/2 Heisenberg chain. In the case of full random matrices, we also derive an analytical expression for the entire evolution of the time-averaged survival probability.
\end{abstract}

\maketitle

\section{Introduction}

Fractal dimensions are tools for probing scaling behavior near quantum phase transitions and for quantifying the complexity of quantum systems~\cite{Wegner1980,Soukoulis1984,Castellani1986,Ketzmerick1992,Huckestein1994,Zhong1995,Yuan2000,Huckestein1995,Huckestein1997,Huckestein1999,Schreiber1991,Evers2000,evers2008,Yang2017,VarmaPRE2017,Deng2019,Xu2020,Jagannathan2021, Torres2015,Luitz2016,Schiulaz2019,Laflorencie2020,Solorzano2021,Wang2021,Sarkar2022,Sierant2022,Bastarrachea2024,Bhakuni2024,Sierant2025,Das2025Arxiv,Hamanaka2025,Varga2000,Mirlin2000statistics,Varga2002,Bogomolny2011,Bermudez2012,Laflorencie2016,Backer2019,Carrera2021,Chen2024,Buijsman2025Arxiv}. They provide insight into the structural properties of quantum states, enabling the distinction between localized and extended phases in both noninteracting~\cite{Wegner1980,Soukoulis1984,Castellani1986,Ketzmerick1992,Huckestein1994,Zhong1995,Yuan2000,Huckestein1995,Huckestein1997,Huckestein1999,Schreiber1991,Evers2000,evers2008,Yang2017,VarmaPRE2017,Deng2019,Xu2020,Jagannathan2021} and interacting quantum models ~\cite{Torres2015,Luitz2016,Schiulaz2019,Laflorencie2020,Solorzano2021,Wang2021,Sarkar2022,Sierant2022,Bastarrachea2024,Bhakuni2024,Sierant2025,Das2025Arxiv,Hamanaka2025}, and revealing universal features at critical points~\cite{Schreiber1991,Kravtsov1997,Kravtsov1999,Kravtsov2000,Macleans2003,Monthus2016,Monthus2017}. Fractal dimensions also offer a means to explore the Hilbert space accessed under time evolution, providing information about quantum dynamics~\cite{Torres2015,Laflorencie2016,Solorzano2021,Wang2021,SierantPRB2022,Sierant2022,Sarkar2022,Turkeshi2023,Yousefjani2023,Hopjan2023,Bhakuni2024}. 

The focus of this work is the correlation fractal dimension $D_2$, which has had a predominant role in studies of localization and multifractality. As any fractal dimension, $D_2$ is rigorously defined in the thermodynamic limit. However, since numerical studies are restricted to finite systems, $D_2$ is extracted from scaling analyses of the inverse participation ratio (IPR) \cite{Izrailev1990}, requiring access to many different system sizes and full exact diagonalization. This procedure is computationally demanding, particularly when dealing with many-body quantum systems, because  their Hilbert space dimension grows exponentially with system size. This makes it challenging to reliably obtain $D_2$, especially in the vicinity of critical points, where fluctuations are strong and convergence is slow~\cite{Solorzano2021}.

We explore an alternative and efficient method for extracting the fractal dimension $D_2$, circumventing the limitations of scaling analyses. Instead of relying on the direct analysis of the structures of the eigenstates, we obtain $D_2$ dynamically from the time-averaged survival probability, that is, the time-averaged fidelity between an initial state and its time-evolved counterpart~\cite{Ketzmerick1992,Huckestein1994,Zhong1995,Yuan2000,CROliveira1999,Evangelou1993,Kawarabayashi1996}. The power-law exponent governing the long-time decay of the time-averaged survival probability coincides with the fractal dimension $D_2$. This dynamical approach gives access to larger system sizes than those achieved via exact diagonalization, thanks to advanced algorithms developed for simulating quantum dynamics~\cite{Karrasch2013,Chanda2020,Strathearn2018,Hu2020,Banuls2017}. 

An additional advantage of extracting $D_2$  from dynamics is that the basis is unambiguously defined, corresponding to the energy eigenbasis. In contrast, usual calculations of fractal dimensions based on the IPR of eigenstates depend on the choice of basis, such as real space or momentum space. Nevertheless, the results of both the eigenstate-based and dynamical approaches depend on the energy window considered. Initial states are selected with energies close to the center of the spectrum, where many-body eigenstates typically exhibit the maximal degree of delocalization permitted by the system's regime. 
Furthermore, averaging over many initial states and disorder realizations reduces fluctuations and potential biases associated with specific state choices.

We compare four quantities: the power-law exponent governing the decay of the survival probability, the power-law exponent of the time-averaged survival probability decay, the fractal dimension $D_2^{\text{IPR}_0}$ obtained from the scaling analysis of the IPR$_0$ of the initial state in the energy eigenbasis~\cite{Huckestein1997,Huckestein1999,Torres2015,Torres2015BJP}, and the fractal dimension $D_2^{\text{box}}$ extracted from a box counting method~\cite{Grassberger1983,Castellani1986,Halsey1986,Pook1991,Schreiber1991,Ketzmerick1992,Janssen1994,Huckestein1995}. We show that the survival probability is not only sensitive to the properties of the eigenstates, but it is also influenced by both spectral and system boundaries. In contrast,  the time-averaged survival probability filters out spectral details and reflects only correlations among the components of quantum states, making it an appropriate quantity for extracting the fractal dimension from dynamics. We also find that the power-law exponent of the time-averaged survival probability shows closer agreement with $D_2^{\text{box}}$ than with $D_2^{\text{IPR}_0}$, particularly near critical points.

The comparison between the power-law exponent $\nu$ of the time-averaged survival probability decay and $D_2^{\text{box}}$ has been previously explored in noninteracting models, such as the Aubry-Andr\'e (Harper) model with a single excitation and quasiperiodic disorder~\cite{Ketzmerick1992,Yuan2000} and Fibonacci models~\cite{Ketzmerick1992,Zhong1995}, where the eigenstates are multifractal despite the absence of disorder. Our focus is instead on interacting systems, where the analyses of the structure of the eigenstates typically focus on $D_2$ obtained from IPR, with some studies including the power-law exponent $\gamma$ of the survival probability decay ~\cite{Torres2015,Zarate2023,Hopjan2023}. We examine all four quantities  ($D_2^{\text{IPR}_0}$, $D_2^{\text{box}}$, $\gamma$, and $\nu$) in both the random matrix models and the interacting systems, and find excellent agreement between $\nu$ and $D_2^{\text{box}}$ in all cases.

Building on the analytical derivation of the entire time evolution of the survival probability  under full random matrices~\cite{santostorres2017aip,Torres2018,Schiulaz2019}, we also derive an analytical expression for the time-averaged survival probability. The survival probability under full random matrices of the Gaussian orthogonal ensemble (GOE)~\cite{MehtaBook} decays with a power-law exponent $\gamma=3$, which does not match the fractal dimension, since $0 \leq D_2 \leq 1$. In contrast, the time-averaged survival probability decays with an exponent $\nu=1$, consistent with the fractal dimension of eigenstates are maximally delocalized as in full random matrices. This demonstrates our claim that the time-averaged survival probability suppresses the influence of spectral properties and captures correlations among the components of a quantum state, making it the most suitable dynamical quantity for extracting the fractal dimension.

The structure of the paper is as follows. In Sec.~\ref{Sec:Quantities}, we introduce the relevant observables and present the two ways to obtain the fractal dimension $D_2$, based on the IPR and on the box-counting method. In Sec.~\ref{Sec:Random}, we benchmark our approach on random matrix ensembles, using first matrices from the GOE, and then ensembles of random matrices that exhibit a localization-delocalization transition, namely the Rosenzweig-Porter (RP) model~\cite{Rosenzweig1960,Khaymovich2021,Venturelli2023,Zhang2023,Buijsman2024} and the power-law banded random matrix (PBRM) model~\cite{Mirlin1996,Mirlin2000statistics,Varga2000,Varga2002,Bogomolny2018,Rao2022,Tomasi2023,Buijsman2025Arxiv,Mendez2012,Carrera2021,DeTomasi2023}.  In Sec.~\ref{Sec:Many}, we extend our analysis to two many-body quantum systems: the Aubry-Andr\'e model with interactions~\cite{Iyer2013,schreiber2015}, and the disordered Heisenberg spin-1/2 chain \cite{SantosEscobar2004,Santos2009JMP,Suntajs2020,Sierant2025}.

\section{Survival Probability and Fractal Dimension}
\label{Sec:Quantities}

This section introduces the  quantities analyzed in this work. On the dynamical side, we focus on the survival probability and the time-averaged survival probability. To directly probe the structure of the eigenstates, we compute the fractal dimension $D_2$ using two approaches: the box-size scaling method and the scaling analysis of the IPR.

\subsection{Survival Probability}
The survival probability gives the probability of finding
the system in the initial state $\ket{\Psi(0)}$ at time $t$. Mathematically it is given by,
\begin{align}
 S_{P}(t) &= \left| \braket{\Psi (0)}{\Psi (t)} \right|^2 = \left| \sum_{n =1}^{N} \left| C_{n}^{(0)} \right|^2  e^{-iE_{n}t} \right|^2 \\
 &= \sum_{n \neq m}^{N} \left| C_{n}^{(0)} \right|^2 \left| C_{m}^{(0)} \right|^2  e^{-i(E_{n} - E_{m})t} + \text{IPR}_0 , \nonumber
\end{align}
where $C_{n}^{(0)}=\braket{\psi_{n}}{\Psi (0)}$ are the coefficients of the initial state projected in the energy eigenbasis $\lbrace|\psi_{n} \rangle \rbrace$ of the system's Hamiltonian, $H|\psi_{n} \rangle = E_{n} |\psi_{n} \rangle$, $N$ is the dimension of the Hilbert space, and $\text{IPR}_{0}$ is the saturation value of $S_P(t)$ in the absence of too many degeneracies. This asymptotic value corresponds to the IPR of the initial state in the energy eigenbasis,
\begin{equation}
    \text{IPR}_{0} = \sum_{n=1}^{N}  \left| C_{n}^{(0)} \right|^4.
    \label{Eq:IPR0}
\end{equation}

At very short times, the survival probability shows a universal quadratic behavior followed by dynamical features that depend on the model and initial state~\cite{torres2014njp,torres2014prab}.
At intermediate times, the survival probability averaged over initial states and disorder realizations often exhibits a power-law decay 
\begin{equation}
\langle S_{P}(t)\rangle
\propto t^{-\gamma} .
\label{Eq:Spgamma}
\end{equation}
In the equation above, $\langle \cdot \rangle$ indicates average and the exponent $\gamma$ carries information about the components of the eigenstates~\cite{chalker1990,tavora2016,tavora2017,Torres2015,Torres2017} and the bounds of the spectrum~\cite{Khalfin1958,fonda1978,urbanowski2009,tavora2016,tavora2017}.

The time-averaged survival probability 
\begin{equation}
    \overline{\langle S_{P}(t) \rangle} = \frac{1}{t} \int_0^{t} \left\langle S_{P}(\tau) \right\rangle d \tau,
    \label{Eq:Ct}
\end{equation}
reduces the temporal fluctuations present in $S_{P}(t)$ and also exhibits a power-law decay, 
\begin{equation}
\overline{\langle S_{P}(t) \rangle}
\propto t^{-\nu} ,
\end{equation}
where $\nu$ does not necessarily coincide with $\gamma$ in Eq.~(\ref{Eq:Spgamma}). In this work, we compare $\gamma$ and $\nu$ with the fractal dimension $D_2$. 

For the time evolution, a total number of $10^4$ samples of initial states and disorder realizations are used for the averages. The initial states are selected so that their energies, $\langle \Psi(0)|H|\Psi(0)\rangle$, lie near the center of the spectrum.

\subsection{Fractal Dimension}
\label{Sec:D2s}

The value of the fractal dimension $D_2$ has information about the structure of a quantum state. The state is fully delocalized (ergodic) when $D_2 = 1$, extended but nonergodic when $0<D_2 <1$, and localized if $D_2 = 0$. 
We describe below two common approaches to extract $D_2$.

(i) One method is based on the scaling analysis of the IPR$_0$ [Eq.~\ref{Eq:IPR0}] with respect to the Hilbert space dimension $N$~\cite{Bauer1990,Evers2000,Mirlin2000,Mildenberger2002,Varga2002}, since 
\begin{equation}
  \text{IPR}_0 \propto N^{-D_2^{\text{IPR}_0}} . 
\end{equation}
The analysis is performed by plotting the average $\langle \ln (\text{IPR}_0) \rangle$ versus $\ln N $, where the negative slope gives an estimate of $D_2^{\text{IPR}_0}$. Using $\langle \ln (\text{IPR}_0) \rangle$ instead of $\ln \langle \text{IPR}_0 \rangle$ reduces statistical fluctuations more effectively. While some studies analyze the scaling of the IPR for eigenstates in a fixed basis, we find that this approach can lead to significant deviations from $\nu$, especially near critical regions. Overall, $\nu$ and quantities that agree with it are more sensitive to multifractality than $D_2$ obtained from eigenstates.

(ii) The other approach to the fractal dimension $D_2$ is the box-counting method~\cite{Grassberger1983,Castellani1986,Halsey1986,Pook1991,Schreiber1991,Ketzmerick1992,Janssen1994,Huckestein1994,Zhong1995,Yuan2000}. It consists of dividing the spectrum into boxes $B_i$ of size $\l$ and computing the probability $\sum_{E_{n} \in B_{i}} |C_{n}^{(0)}|^2 $, which is obtained with the components of the initial state corresponding to the eigenstates with energy $E_{n}$ in $B_{i}$. The total probability is the sum over the total number $N_B(l)$ of boxes of size $0<l\leq1$, 
\begin{equation}
   P(l) = \sum_{i=1}^{N_B(l)} \left( \sum_{E_{n} \in B_{i}} |C_{n}^{(0)}|^2 \right)^{2}  .
\end{equation}
The fractal dimension $D_2^{\text{box}}$ is extracted from scaling with respect to the box size $l$,
\begin{equation}
    P(l) \propto l^{D_2^{\text{box}}} . 
    \label{eq:gamma_K}
\end{equation}
Above, $P(l)$ is the probability that the difference between the energies of any two eigenstates from the spectral decomposition of $|\Psi(0)\rangle$ is less than $l$. We calculate $P(l)$ by normalizing the spectrum length to $l=1$, so that a box of size $l=1$ includes all components $|C_{n}^{(0)}|^2$ and $P(1)=1$. The scaling analysis is performed for $l \ll 1$.

The fractal dimension $D_2^{\text{box}}$ probes the distribution of the components of $|\Psi(0)\rangle$ in energy at different scales $l$, being a more local measure than $D_2^{\text{IPR}}$. While $D_2^{\text{box}}$ can, in principle, be computed for even a single state,  $D_2^{\text{IPR}}$ often uses ensemble averages. Furthermore, $D_2^{\text{box}}$ is directly related to the exponent of the power-law decay of $\overline{\langle S_P(t) \rangle}$~\cite{Ketzmerick1992}. 

In this work, the results for \( D_2^{\mathrm{IPR_0}} \) obtained from scaling analyses are based on averages over disorder realizations and initial states. All averages consider $10^4$ samples, as in the analysis for the power-law exponents.

\section{Random matrices} 
\label{Sec:Random}

This section compares the power-law decay exponents of $\langle S_P(t) \rangle$ and $\overline{\langle S_P(t) \rangle}$ with $D_2^{\text{IPR}_0}$ and $D_2^{\text{box}}$ for the GOE, RP, and PBRM models.

\subsection{GOE matrices}
The GOE consists of real and symmetric ${N}\times{N}$  matrices completely filled with random numbers from a Gaussian distribution with mean zero and variance given by
\begin{equation}
\label{eq:HamGOE}
\left\langle H_{ij}^2\right\rangle=
\begin{cases}
    1/2,\quad\text{for}\quad i\neq j ,\\
    1,\quad \hspace{.35cm} \text{for}\quad i=j.
\end{cases}    
\end{equation}
The eigenvalues of these matrices are highly correlated~\cite{MehtaBook} and the eigenstates are normalized random vectors. 

We consider initial states corresponding to the diagonal part of the random matrix. Since the eigenstates are normalized random vectors, the components of these initial state, $C_{n}^{(0)}$, are random numbers from a Gaussian distribution with $\langle C_{n}^{(0)} \rangle \sim 0$, $\langle C_{n}^{(0)} \rangle^2 \sim 1/N$, and $\langle C_{n}^{(0)} \rangle^4 \sim 3/N^2$, so  $\langle \text{IPR}_{0} \rangle \sim 3/N $ and $D_2^{\text{IPR}_0} \sim 1$. 

Similarly, the box-counting method gives $D_2^{\text{box}} \sim 1$, since for GOE matrices, we have approximately $lN$ eigenstates with energy $E_n$ in a box of size $l$. This implies that $\sum_{E_{n} \in B_{i}} |C_{n}^{(0)}|^2 \sim l$ and $P(l) \sim l$.

Using GOE matrices, it is possible to derive an analytical expression for the survival probability~\cite{Torres2018,Schiulaz2019} 
\begin{equation}
\label{eq:SP_Analytical}
     \left\langle S_{P}(t) \right\rangle = \frac{1 - \overline{S_{P}}}{{N}-1} \left[ {N} \frac{{\cal{J}}_1^{2} (2 \Gamma t)}{(\Gamma t)^2} - b_2 \left( \frac{\Gamma t}{2 {N}} \right) \right] + \overline{S_{P}} ,
\end{equation}
where  $\Gamma = \sqrt{{N}/2}$ is the width of the energy distribution of the initial state, ${\cal{J}}_1$ is the Bessel function of first kind, $\overline{S_{P}} = 3/({N }+2)$ \cite{alhassid1992,Torres2016entropy} and $b_2 (t)$ is the two-level form factor,
\begin{eqnarray}
\label{eq:b2_GOE}
b_{2}(t) &=& \Theta(1-t) [1-2t + t\text{ln}(2t+1)] \nonumber \\ 
&+& \Theta(t-1) \{ t\text{ln}[ (2t+1)/(2t-1)]-1 \}
\end{eqnarray}
where $\Theta (t)$ is the Heaviside step function. The first term in Eq.~(\ref{eq:SP_Analytical}), involving the Bessel function, controls the short- and intermediate-time decay of the survival probability, the second term, with the $b_2(t)$ function, governs the long-time behavior up to the saturation at $\overline{S_{P}}$.

\begin{figure}[h]
    \centering
\includegraphics[width=.95\linewidth]{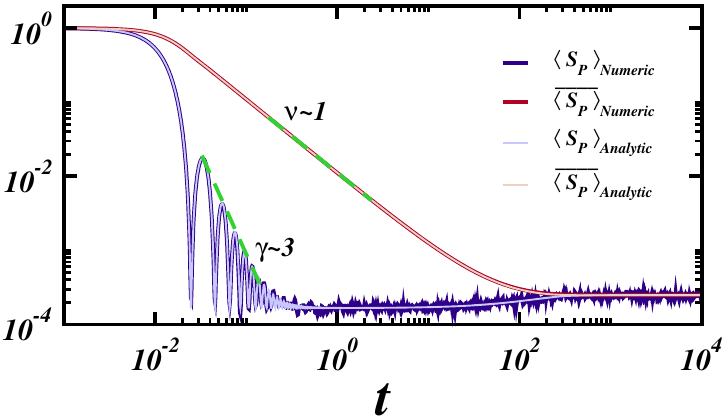}
    \caption{Analytical expression for the survival probability $\langle S_{P}(t) \rangle$ in Eq.~(\ref{eq:SP_Analytical}) (light blue), numerical results (dark blue), analytical expression for the time-averaged survival probability $\overline{\langle S_P(t) \rangle}$ in Eq.~(\ref{eq:TimeAvSp_Analytical_GOE}) (dark red), and numerical results (light red) for matrices from the GOE; $N = 12\,000$.}
    \label{fig:RMT}
\end{figure}

Figure~\ref{fig:RMT} compares the analytical expression in Eq.~(\ref{eq:SP_Analytical}) (light blue line) with the corresponding numerical result (dark blue line). The oscillations  observed in both curves originate from the term with the Bessel function in the analytical expression and exhibit a power-law decay $\propto t^{-3}$. The Bessel function arises from the square of the Fourier transform of the energy distribution of the initial state, which follows a semicircular form~\cite{torres2014njp}. The power-law exponent $\gamma=3$ is a consequence of the sharp edges of the semicircular energy distribution, being clearly unrelated with the fractal dimension $D_2^{\text{IPR}_0}=1$. The exponent $\gamma=3$ reflects spectral properties rather than the structure of the eigenstates.

At long times, $\langle S_P(t)\rangle$ shows a dip below the saturation value known as the correlation hole~\cite{alhassid1992,Leviandier1986,Torres2017philo,Das2025}. It is characterized by the $b_2(t)$ function and emerges only in systems with level repulsion.

The decay behavior of the survival probability changes when time averaging is introduced. For the time-averaged survival probability $ \overline{\langle S_{P}(t) \rangle}$, the power-law decay exponent becomes $\nu=1$ in contrast to the faster decay of the survival probability. By performing the integration in Eq.~(\ref{Eq:Ct}) (see Appendix \ref{Appendix A}), we obtain the analytical expression
\begin{equation}
\label{eq:TimeAvSp_Analytical_GOE}
    \overline{\langle S_{P}(t) \rangle} = \frac{\overline{S_{P}}}{3}\left[4 N{\cal{I}}_{1}(2\Gamma t) - {\cal{I}}_{2}\left(\frac{\Gamma t}{2N}\right)\right] + \overline{S_{P}}.
\end{equation}
For times, $\Gamma t \gg 1$, we reach the asymptotic results
\begin{align}
\label{eqn: TimeAvSpGOEAsymptitic}
    {\cal{I}}_{1}(2\Gamma t) &\approx \frac{2}{3\pi \Gamma t} + O(t^{-2}) ,\\
    {\cal{I}}
    \label{eqn: TimeAvSpGOEAsymptitic02}_{2}\left(\frac{\Gamma t}{2N}\right) &\approx \frac{N}{2\Gamma t} + O(t^{-2}),
\end{align}
leading to the large-time behavior
\begin{equation}
\label{eq:SatTimeAvSp_Analytical_GOE}
    \overline{\langle S_{P}(t\to\infty) \rangle} \approx \frac{8}{9\pi}\frac{N}{\Gamma t}\overline{S_{P}} - \frac{1}{6}\frac{N}{\Gamma t}\overline{S_{P}} + \overline{S_{P}} + O(t^{-2}).
\end{equation}
This result confirms that $\overline{\langle S_P(t) \rangle }\propto t^{-1}$. The absence of Bessel-function-induced oscillations reflects the fact that information about the spectrum has been averaged out. The decay of the time-averaged survival probability is governed by the structure of the initial states and the power-law decay exponent $\nu$ agrees with the fractal dimension $D_2^{\text{box}}$. When the components of the initial states and of the eigenstates are uncorrelated random variables, $\nu=D_2^{\text{box}}=D_2^{\text{IPR}_0}=1$.

Figure~\ref{fig:RMT} shows that the analytical result for the time-averaged survival probability in Eq.~\eqref{eq:TimeAvSp_Analytical_GOE} (dark red line) is in excellent agreement with the numerical result (light red line). The power-law decay persists up to saturation, with no indication of a correlation hole, and the exponent $\nu=1$ coincides with the fractal dimension.

\subsection{Rosenzweig-Porter matrices}

The RP model was introduced in the context of nuclear physics~\cite{Rosenzweig1960} to study the transition between regular and chaotic spectra in complex nuclei. The ensemble consists of random matrices with entries taken from a Gaussian distribution with mean zero and variance
\begin{equation}\label{eq:varPRBM}
\left\langle H_{ij}^2\right\rangle=\begin{cases}
1, & i=j\\
    \frac{1}{2} {N}^{-\alpha}, & i\neq j, 
    \end{cases}
\end{equation}
where $N$ is the dimension of the matrix and $\alpha$ is the control parameter. 
The ensemble presents different phases depending on the value of $\alpha$: an ergodic phase when $\alpha< 1$, a nonergodic extended phase with extended nonergodic states when $1< \alpha <2$, and a localized phase for $\alpha > 2$ ~\cite{Pino2019,Von2019,Zhang2023,Nosov2019}. We recover the GOE when $\alpha=0$ and a diagonal random matrix is obtained for $\alpha \rightarrow \infty$.

We begin by comparing the power-law exponents $\gamma$ and $\nu$ obtained from the decay of the survival probability $\langle S_{P}(t) \rangle$ and the time-averaged survival probability $\overline{\langle S_{P}(t) \rangle}$, respectively, across different values of $\alpha$. Next, we illustrate the procedure for computing the fractal dimensions $D_2^{\text{IPR}_0}$ and $D_2^{\text{box}}$ through explicit examples. Finally, we compare all relevant quantities:  $\gamma$, $\nu$  $D_2^{\text{IPR}_0}$, and $D_2^{\text{box}}$. Our analysis shows that $\nu$ from $\overline{\langle S_{P}(t) \rangle}$ and $D_2^{\text{box}}$ are in agreement and offer the most reliable indicators for identifying the system’s regime, whether delocalized, critical, or localized.

\begin{figure}[h]
    \centering
    \includegraphics[width=1\linewidth]{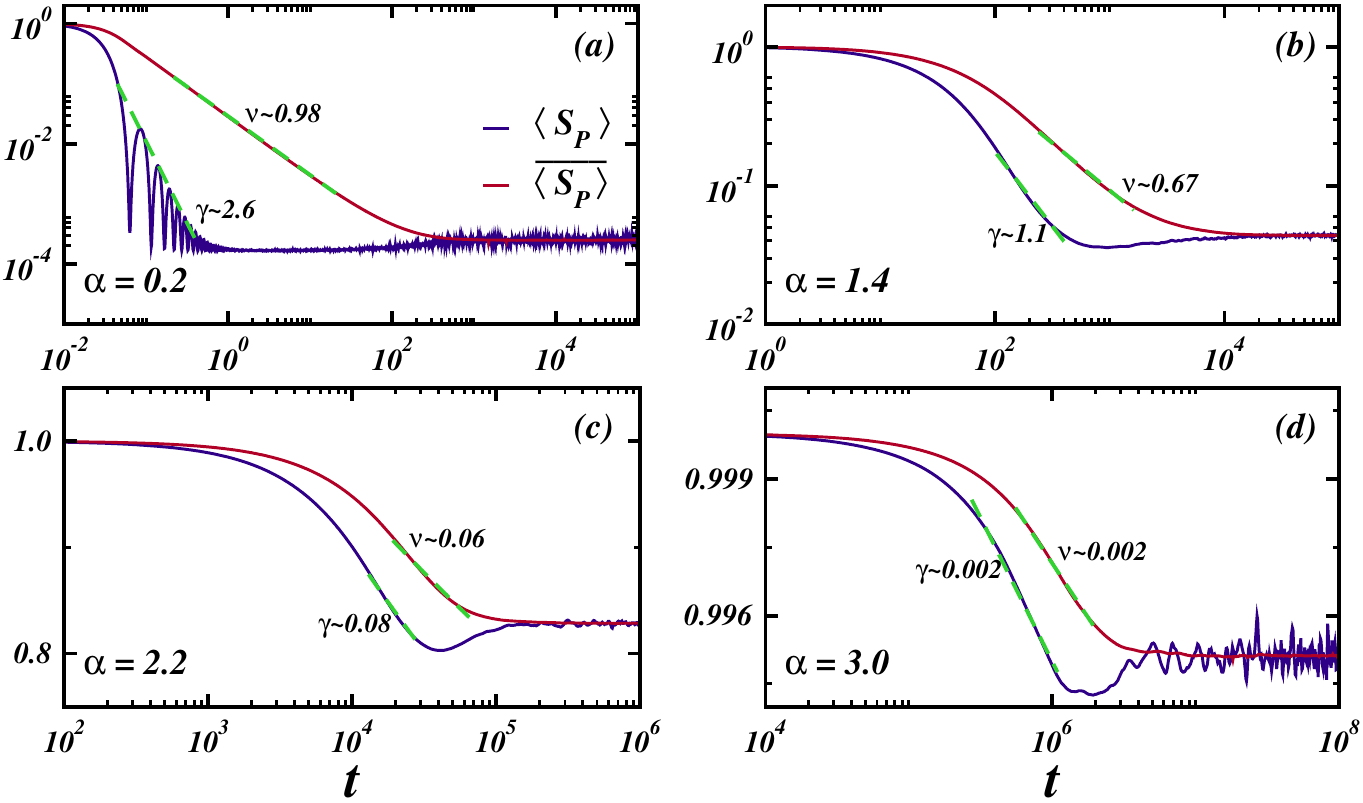}
    \caption{Survival probability $\langle S_P(t) \rangle$ (dark blue) and time-averaged survival probability $\overline{\langle S_P(t) \rangle}$  (dark red) for the Rosenzweig-Porter model. The power-law decays $\propto t^{-\gamma}$ and $\propto t^{-\nu}$ are indicated with green dashed lines and the values of $\gamma$ and $\nu$ are shown in each panel. The time intervals chosen for the fittings are the same for both exponents. The values of the control parameter $\alpha$ are also indicated; $N=12\,000$.  }
    \label{fig:CT_SP_RPE}
\end{figure}

In Fig.~\ref{fig:CT_SP_RPE}(a), where $\alpha=0.1$, the behaviors of $\langle S_{P}(t) \rangle$ and $\overline{\langle S_{P}(t) \rangle}$ are similar to those for the GOE in Fig.~\ref{fig:RMT}, with $\gamma>D_2^{\text{IPR}_0} \sim 1$ for $\langle S_{P}(t) \rangle$ and $\nu=D_2^\text{box} \sim 1$ for $\overline{\langle S_{P}(t) \rangle}$. As $\alpha$ increases from the delocalized regime [Fig.~\ref{fig:CT_SP_RPE}(a)] to the localized regime in Fig.~\ref{fig:CT_SP_RPE}(d), both power-law exponents decrease, with $\gamma$ remaining larger than $\nu$ throughout. The behavior of $\langle S_P(t) \rangle$ changes significantly from the delocalized [Fig.~\ref{fig:CT_SP_RPE}(a)] to the intermediate regime [Fig.~\ref{fig:CT_SP_RPE}(c)]. In the first case, $\langle S_P(t) \rangle$ exhibits oscillations during its power-law decay and a later correlation hole, while in the intermediate regime, the  oscillations vanish and the correlation hole gradually disappears.

\begin{figure}[h]
    \centering
    \includegraphics[width=0.75\linewidth]{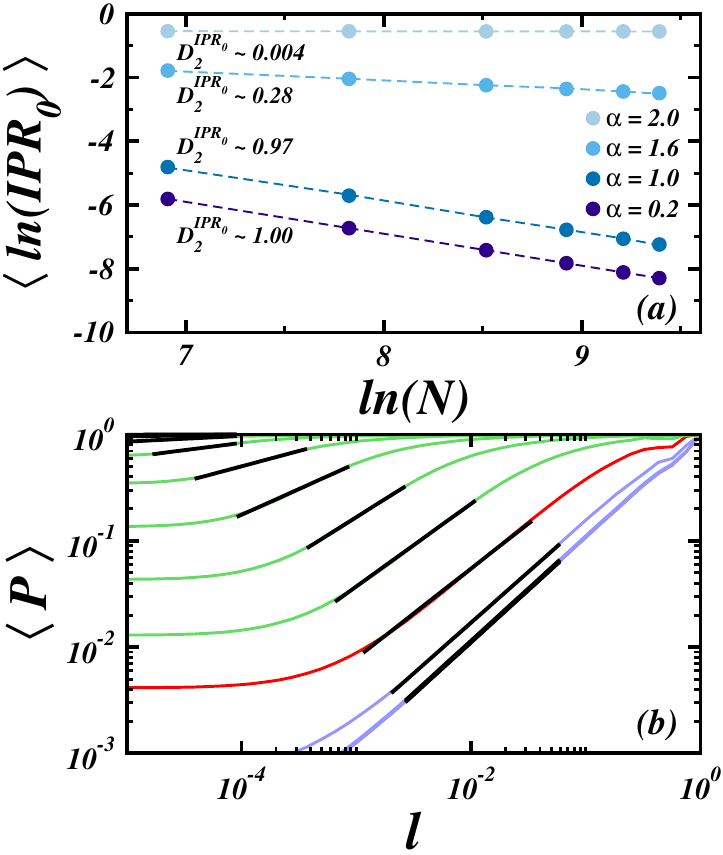}
    \caption{(a) Scaling analysis of $\langle \ln (\text{IPR}_0)\rangle$ to extract $D_2^{\text{IPR}_{0}}$ and (b) scaling analysis of $\langle P(l)\rangle$ from Eq.~(\ref{eq:gamma_K}) to obtain $D_2^{\text{box}}$ for the Rosenzweig–Porter model.  The values of $\alpha$ are indicated in (a). In (b), the blue lines are for $\alpha<1$, the red curve corresponds to $\alpha=1$, the green lines are for $1<\alpha$, and the black lines are fitting curves; $N=12\,000$.  }
\label{fig:RPM_gamma_D2_tilde}
\end{figure}

In Fig.~\ref{fig:RPM_gamma_D2_tilde}, we illustrate how the fractal dimension is obtained from the IPR  [Fig.~\ref{fig:RPM_gamma_D2_tilde}(a)] and from the box-counting method [Fig.~\ref{fig:RPM_gamma_D2_tilde}(b)] for some values of the control parameter $\alpha$. In Fig.~\ref{fig:RPM_gamma_D2_tilde}(a), we fit the data for $\langle\ln (\text{IPR}_0) \rangle$ vs $\ln N$ with a straight line, and the slope corresponds to $D_2^{\text{IPR}_{0}}$. In Fig.~\ref{fig:RPM_gamma_D2_tilde}(b), the curves for $P(l)$ are nearly flat for small $l$ and they saturate for large $l$. We obtain $D_2^{\text{box}}$ from the interval of the log-log plot where the curves are linear.

\begin{figure}[h]
    \centering
    \includegraphics[width=0.75\linewidth]{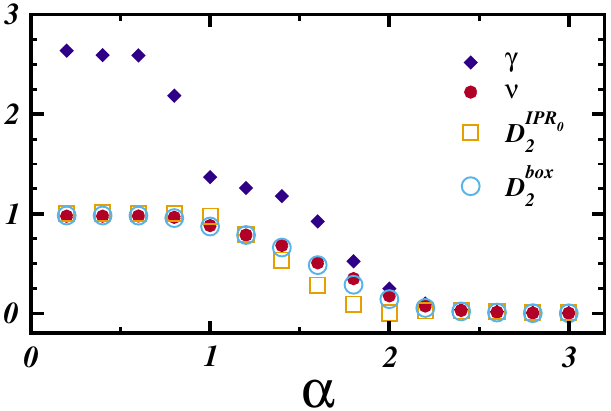}
    \caption{Power-law exponents $\gamma$ and $\nu$, and fractal dimensions $D_{2}^{\text{IPR}_0}$ and $D_{2}^{\text{box}}$ as a function of the control parameter $\alpha$ for the Rosenzweig-Porter model.   }
\label{fig:D2_spectral_D2_gammaSP_delta_Ct_RPM}
\end{figure}

Figure~\ref{fig:D2_spectral_D2_gammaSP_delta_Ct_RPM} compares all four quantities: $\gamma$, $\nu$, $D_2^{\text{IPR}_{0}}$, and $D_2^{\text{box}}$. There is excellent agreement between the exponent $\nu$ of the time-average survival probability and $D_2^{\text{box}}$ for all values of $\alpha$. The exponent $\nu$ and the fractal dimension $D_2^{\text{box}}$ also align well with $D_2^{\text{IPR}_{0}}$ in the ergodic and localized regimes, but not so well in the extended nonergodic phase. The exponent $\gamma$, associated with the survival probability, deviates from all three quantities throughout the delocalized and intermediate regimes, only approaching agreement in the localized phase ($\alpha>2$).  In the delocalized phase, $\gamma>1$ and it approaches $\gamma=3$ for $\alpha \rightarrow 0$, as in the GOE model.

It is interesting that the four quantities distinguish the three regions: $\alpha<1$, $1<\alpha<2$, and $\alpha>2$. They decay as $\alpha$ increases in the nonergodic extended phase and are nearly constant in the ergodic and localized regimes.

\subsection{Power-law banded matrices}
\label{Sec:PBRM}

The PBRM model corresponds to an ensemble of real and symmetric $N\times N$  matrices filled with real Gaussian random numbers whose mean is zero and the variance of its elements is given by 
\begin{equation}\label{eq:varPRBM}
\left\langle H_{ij}^2\right\rangle=\begin{cases}
1, & i=j\\
    \left[1+ \left|\left( i-j \right) /\beta \right|^{2 \alpha} \right]^{-1}, & i\neq j, 
    \end{cases}
\end{equation}
where $\beta$ is the bandwidth and the control parameter $\alpha$ can be varied from the delocalized phase ($\alpha< 1$) to the localized phase ($\alpha>1$), with $\alpha=1$ being the critical point~\cite{Mirlin1996,Kravtsov1997,Mirlin2000statistics,Varga2000,evers2008,Rao2022} We recover the GOE matrix when $\alpha \rightarrow 0$ and the tight-biding model with nearest-neighbor random coupling (tridiagonal matrix) when $\alpha \rightarrow \infty$.
Physically, the PBRM model can be interpreted as describing a particle in a one-dimensional disordered system with long-range hopping that decays as a power law~\cite{Mirlin1996,Evers2000,Mirlin2000}. 

The figures used to extract $\gamma$, $\nu$, $D_2^{\text{IPR}_{0}}$, and $D_2^{\text{box}}$ are provided in the Appendix~\ref{Appendix B}. In Fig.~\ref{fig:D2_spectral_D2_gammaSP_delta_Ct_PBRM}(a), we compare the behavior of these four quantities as a function of the control parameter $\alpha$ for fixed $\beta=1$. The results are similar to those for the RP model, in the sense that there is excellent agreement between $\nu$ and $D_2^{\text{box}}$, and  reasonable agreement also with $D_2^{\text{IPR}_{0}}$,  for any value of $\alpha$, while $\gamma$ only approaches $\nu$ in the localized regime ($\alpha>1$). 
In the delocalized region, $\gamma$ decays abruptly as $\alpha$ increases.

\begin{figure}[h]
    \centering
    \includegraphics[width=0.8\linewidth]{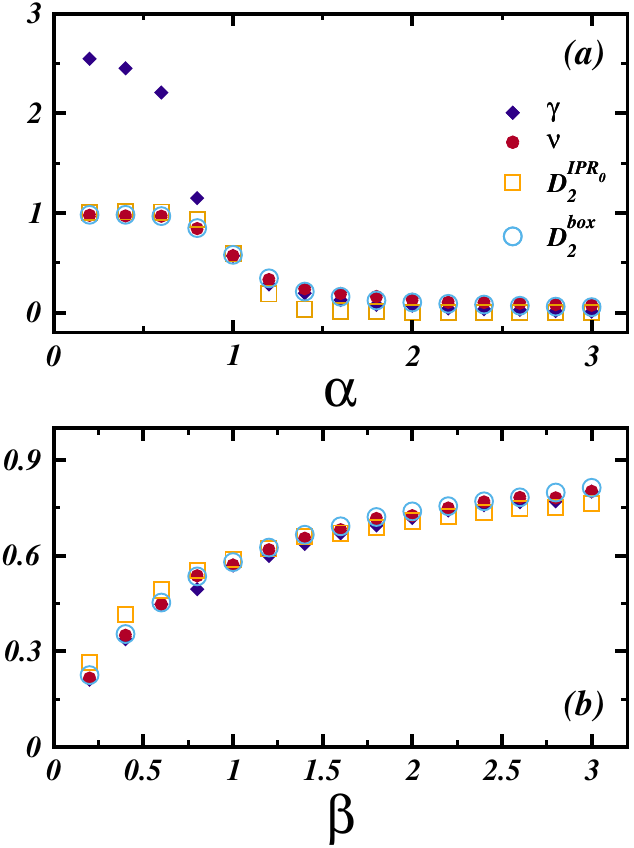}
    \caption{Study of fractality for the power-law banded random matrix model. Power-law exponents $\gamma$ and $\nu$, obtained for $N=12\,000$, and fractal dimensions $D_{2}^{\text{IPR}_0}$ and $D_{2}^{\text{box}}$ 
    as a function of (a) the control parameter $\alpha$ for fixed $\beta=1$ and (b) the bandwidth $\beta$ for the critical point $\alpha=1$. 
    }
\label{fig:D2_spectral_D2_gammaSP_delta_Ct_PBRM}
\end{figure}

The largest fluctuations in the structures of the eigenstates happen at critical points, where the eigenstates are multifractal. This motivates the analysis in Fig.~\ref{fig:D2_spectral_D2_gammaSP_delta_Ct_PBRM}(b), where we fix $\alpha=1$ and compare the behavior of the four quantities, $\gamma$, $\nu$, $D_2^{\text{IPR}_{0}}$, and $D_2^{\text{box}}$, as a function of the bandwidth $\beta$. By increasing $\beta$, we can modify the degree of multifractality of the states, going from  strong multifractality ($\beta \ll 1$) to weak multifractality ($\beta \gg 1$). As before, we find excellent agreement between $\nu$ and $D_2^{\text{box}}$, and good agreement with  $D_2^{\text{IPR}_{0}}$. Interestingly, the power-law exponent $\gamma$ is also very similar to $\nu$, indicating that at the critical point, the power-law decay of $\langle S_P(t) \rangle$ and $\overline{\langle S_P(t) \rangle}$ are comparable, as indeed seen in Fig.~\ref{fig:09} provided in the Appendix~\ref{Appendix B}.

\section{Interacting Many-Body Quantum Models}
\label{Sec:Many}

The spin-1/2 Hamiltonian for the two one-dimensional (1D) physical models with nearest-neighbor couplings that we consider is given by
\begin{equation}
H=\frac{J}{4} \sum_{k=1}^{L-1} \left[ \sigma_{k}^{x} \sigma_{k+1}^{x} + \sigma_{k}^{y} \sigma_{k+1}^{y} + \sigma_{k}^{z} \sigma_{k+1}^{z} \right] + \frac{1}{2} \sum_{k=1}^{L} h_{k} \sigma_{k}^{z} ,
\end{equation}
where $\sigma_{k}^{x,y,z}$ are Pauli matrices that act on the particle at site $k$, the coupling strength is set to $J=1$, and $h_k$ corresponds to Zeeman splittings. The two models differ by the kind of onsite disorder that is chosen. 

$\bullet$ We have the interacting Aubry-André model when $h_k=h \cos\left(2 \pi \beta k + \phi \right)$, where $\beta = \left(1 + \sqrt{2} \right)/2$ is the inverse golden ratio, $\phi$ is a random phase between $0$ and $2 \pi$, and $h$ is the amplitude of the disorder strength. This model transitions from a delocalized phase when $h<0.7$ to a localized regime when $h>1.7$, with an extended nonergodic phase existing for $0.7<h<1.7$ \cite{Xu2019,Shenglong2019}.

$\bullet$ We have the disordered spin-1/2 Heisenberg model, when $h_{k}$ are uncorrelated random numbers taken from a flat distribution with $[-h,h]$ and $h$ being the disorder strength. For $h=0$, the model is integrable. As $h$ increases from zero, the system becomes chaotic. Several numerical studies indicate that this transition should occur for an infinitesimally small value of $h$ as $L \rightarrow \infty$. As the disorder is further increased, beyond the coupling strength, some studies suggest that the model transitions from a delocalized phase to a many-body localized regime,  although there is no consensus on whether this holds~\cite{Luitz2015,Serbyn2015,Suntajs2020,Colbois2024,Niedda2024arxiv}.

We construct the Hamiltonian matrix for both models in the largest subspace of the Hilbert space for which the total $z$-magnetization $\frac{1}{2}\sum_{k=1}^{L} \sigma_{k}^{z} =0$ and the dimension $N=L!/((L/2)!)^2$. The figures used to extract $\gamma$, $\nu$, $D_2^{\text{IPR}_{0}}$, and $D_2^{\text{box}}$ for both models are provided in the appendix. Below, we compare the results as a function of the potential strength $h.$

\subsection{Interacting Aubry-André model}

\begin{figure}[h]
    \centering
\includegraphics[width=0.85\linewidth]{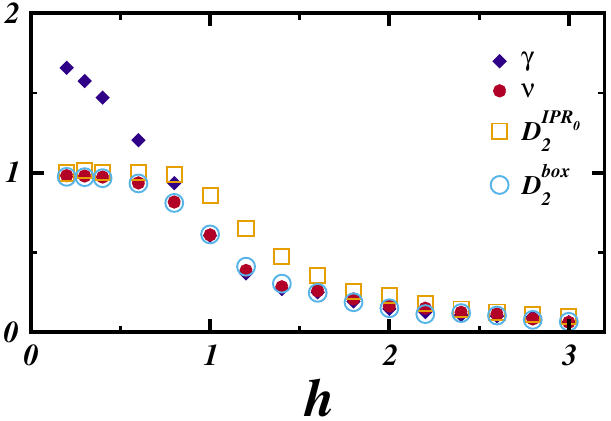}
    \caption{Power-law exponents $\gamma$ and $\nu$ and fractal dimensions $D_{2}^{\text{IPR}_0}$ and $D_{2}^{\text{box}}$ as a function of the potential strength $h$ for the interacting Aubry-André model with largest system size $L=16$ ($N=12\,870$). 
    }  \label{fig:AAM_Int_D2_spectral_D2_gamaSP_deltaCt_Diag}
\end{figure}

Figure~\ref{fig:AAM_Int_D2_spectral_D2_gamaSP_deltaCt_Diag}, obtained for the interacting  Aubry-André model, shows excellent agreement between the power-law decay exponent $\nu$ of $\overline{\langle S_{P}(t)\rangle}$ and the fractal dimension $D_2^{\text{box}}$, supporting our claim that these two quantities coincide in interacting many-body quantum systems. We also observe that in the intermediate regime ($0.7<h<1.7$), the fractal dimension $D_2^{\text{IPR}_0}$ now deviates more from $\nu$ and $D_2^{\text{box}}$ than what was seen for the RP and PBRM models. Even though $D_2^{\text{box}}$ and $D_2^{\text{IPR}_0}$ are both derived from the components $C_{n}^{(0)}$ of the initial state, they rely on different approaches: $D_2^{\text{IPR}_0}$ is based solely on the sum of $|C_n^{(0)}|^4$, while the box-counting method involves sums of components in windows of energies. The intermediate phase of the Aubry-Andr\'e model makes the difference between these two constructions more evident. Furthermore, the faster decay of $D_2^{\text{box}}$ and $\nu$ with increasing $h$ emphasizes their stronger sensitivity to multifractality compared to $D_2^{\text{IPR}_0}$.

The results for $\gamma$ in Fig.~\ref{fig:AAM_Int_D2_spectral_D2_gamaSP_deltaCt_Diag} show that apart from the ergodic region, there is very good agreement with $\nu$ for $h>0.7$, that is, in the intermediate and localized regions. Deep in the chaotic region, $\gamma$ is expected to get close to 2, consistent with the bounds of the energy distribution of the initial state~\cite{tavora2016,tavora2017}, which for physical many-body quantum systems is Gaussian. Deviations from $\gamma=2$ are caused by finite-size effects~\cite{Torres2019}. As $h$ increases and the system approaches the intermediate region, $\gamma$ decays abruptly. In the localized region, all four quantities are very close to zero.

\subsection{Disordered spin-1/2 Heisenberg model}

\begin{figure}[h]
    \centering
\includegraphics[width=0.85\linewidth]{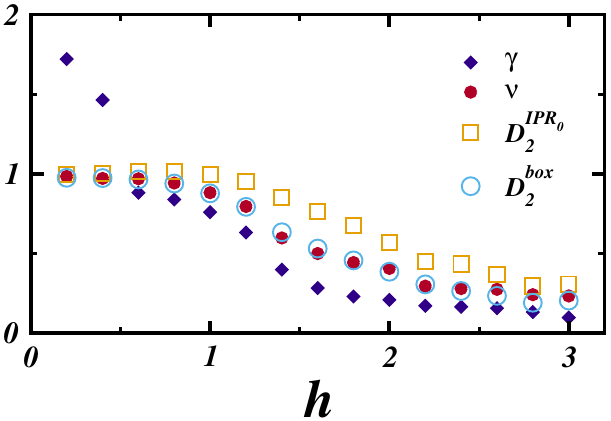}
    \caption{Power-law exponents $\gamma$ and $\nu$ and fractal dimensions $D_{2}^{\text{IPR}_0}$ and $D_{2}^{\text{box}}$ as a function of the potential strength $h$ for the spin-1/2 Heisenberg model with largest system size $L=16$ ($N=12\,870$). 
    }   \label{fig:XXZ_Int_D2_spectral_D2_gamaSP_deltaCt}
\end{figure}

Figure~\ref{fig:XXZ_Int_D2_spectral_D2_gamaSP_deltaCt} shows $\gamma$, $\nu$, $D_2^{\text{IPR}_{0}}$, and $D_2^{\text{box}}$ as a function of the disorder strength $h$ for the Heisenberg model. Once again we confirm that $\nu \approx D_2^{\text{box}}$ for interacting many-body quantum systems. The figures also reiterates that, similarly to the Aubry-Andr\'e model, $\gamma \rightarrow 2$ deep in the chaotic region, decaying as the disorder increases. Furthermore, as before, $\nu$ and $D_2^{box}$ decay faster than $D_2^{\text{IPR}_{0}}$ as $h$ increases above the coupling strength. 

Contrary to the Aubry-Andr\'e model, $\gamma$ in Fig.~\ref{fig:XXZ_Int_D2_spectral_D2_gamaSP_deltaCt}  deviates from $\nu$ for all values of the disorder strength that are shown for $h>J$. At large disorder strengths, all four quantities,  $\nu$, $D_2^{\text{box}}$,  $D_2^{\text{IPR}_0}$, and $\gamma$ exhibit fluctuations rather than a monotonic decay to zero. This behavior reflects the ongoing uncertainty surrounding the existence of a true many-body localized phase for this model.

\section{Conclusions}

This work establishes the time-averaged survival probability as a powerful and practical tool for extracting the fractal dimension $D_2$ in many-body quantum systems. Contrary to conventional methods that require finite-size scaling analysis of measures of delocalization, such as the inverse participation ratio, our approach resorts to dynamics to directly probe the structure of the states. 

Using  ensembles of Gaussian orthogonal, Rosenzweig-Porter, and power-law banded random matrices, and interacting many-body quantum systems described by the Aubry-Andr\'e and disordered Heisenberg models, we demonstrated that the exponent $\nu$ of the power-law decay of the time-averaged survival probability consistently coincides with the fractal dimension $D_2^{\text{box}}$ evaluated using the box-counting method. The agreement between the two quantities, $\nu \approx D_2^{\text{box}}$, holds even in regimes where the survival probability decay exponent $\gamma$ or fractal dimensions computed from inverse participation ratios (e.g. $D_2^{\text{IPR}_0}$)  fail to match among them and with $\nu$. This happens, for example, deep in the chaotic regime, where $\gamma$ is larger than $\nu$, and in extended nonergodic phases, where $D_2^{\text{IPR}_0}$ is usually larger than $\nu$.

In short, the time-averaged survival probability offers a robust and scalable approach to identifying extended nonergodic phases and multifractal behavior in complex quantum systems. It is a promising method for numerical studies, where the direct analysis of the structure of the states is restricted to small system sizes, and for experimental platforms, as it avoids state tomography.

\begin{figure*}[ht]
    \centering
    I. Aubry-Andr\'e Model \hspace{4.35cm} II. Heisenberg Model\\
\includegraphics[width=0.4\linewidth]{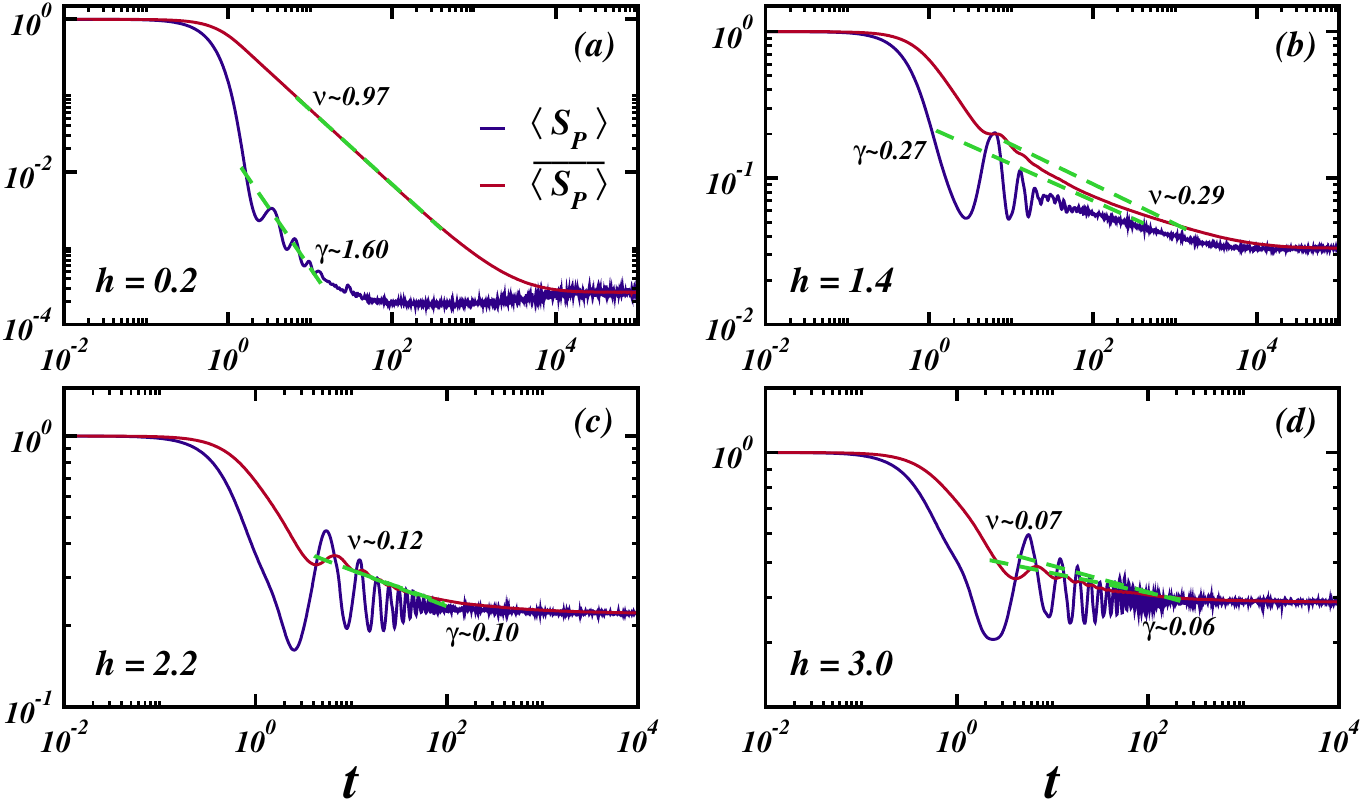} \hspace{0.2 cm}
\includegraphics[width=0.4\linewidth]{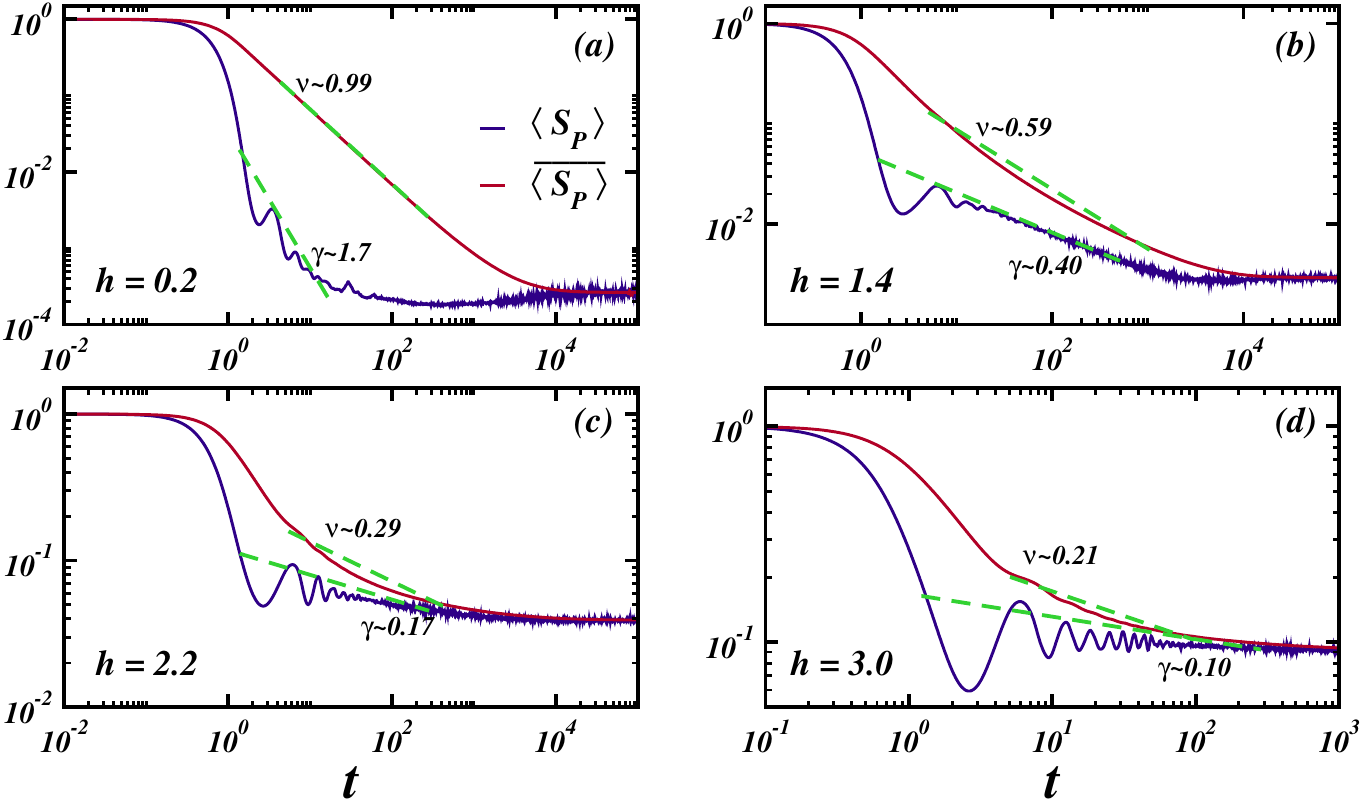} \\
\includegraphics[width=0.3\linewidth]{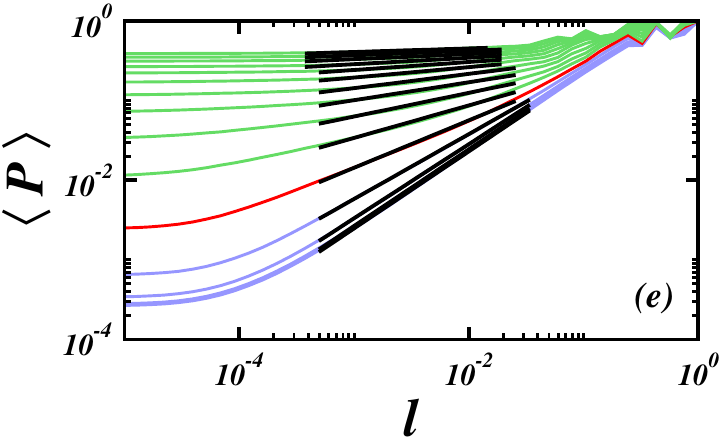} \hspace{2. cm}\includegraphics[width=0.3\linewidth]{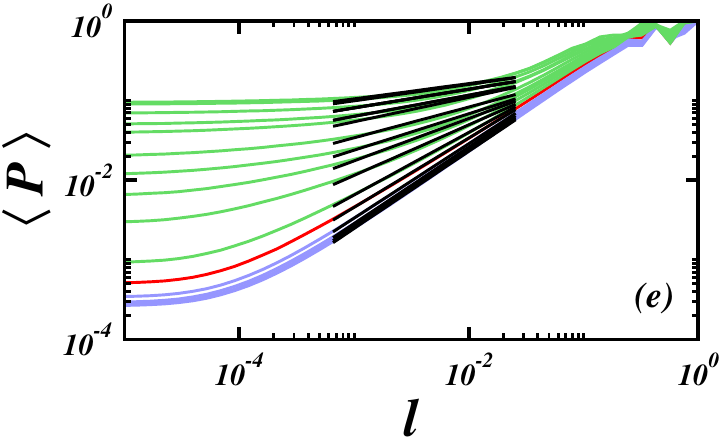}
    \caption{Interacting Aubry-Andr\'e model (left panels) and disordered Heisenberg model (right panels) for $L=16$ ($N=12\,870$). (a)-(d): Survival probability $\langle S_P(t)\rangle$ (dark blue) and time-averaged survival probability $\overline{\langle S_P(t)\rangle}$ (dark red). The power-law decays $\propto t^{-\gamma}$ and $\propto t^{-\nu}$ are indicated with green dashed lines and the values of $\gamma$ and $\nu$ are shown in each panel. The time intervals chosen for the fittings are the same for both exponents. The values of the disordered strength $h$ are also indicated.
    (e): Scaling analysis of $\langle P(l)\rangle$ from Eq.~(\ref{eq:gamma_K}) to obtain $D_2^{\text{box}}$. The blue lines are for $h<1$, the red curve corresponds to $h=1$, the green lines are for $h>1$, and the black lines are fitting curves.}
    \label{fig:08}
\end{figure*}

\section{Acknowledgments}
D. A. Z.-H. and E .J. T.-H. are  grateful to Secihti M{\'e}xico for financial support under project No. CF-2023-I-1748 and to VIEP-BUAP under project No. 100524481-VIEP2024.
I.V.-F. and L.F.S. thank start-up funding from the University of Connecticut.

\appendix
\section{Derivation of the time-averaged survival probability for GOE random matrices}
{\label{Appendix A}}

The time-averaged survival probability is defined as
\begin{equation}
\label{eqn: Def_TimeAvSp}
    \overline{\langle S_{P}(t) \rangle} = \frac{1}{t} \int_0^{t} \left\langle S_{P} (t') \right\rangle d t'
\end{equation}
For the GOE, 
\begin{equation}
\label{eqn: AnalyticSpGOE}
    \left\langle S_{P} (t) \right\rangle = \frac{4}{3}\overline{S_{P}} N\frac{{\cal{J}}_{1}^{2}(2\Gamma t)}{(2\Gamma t)^{2}} - \frac{1}{3}\overline{S_{P}} b_{2}\left(\frac{\Gamma t}{2N}\right) + \overline{S_{P}} ,
\end{equation}
where $\Gamma = \sqrt{N/2}$, and
\begin{equation}
\label{eqn: b2GOE}
b_{2}(t)= f(t)\Theta(1-t) + g(t)\Theta(t-1) ,
\end{equation}
with
\begin{align*}
    f(t) &= 1-2t + t\text{ln}(2t+1),\\
    g(t) &= t\text{ln}\left( \frac{2t+1}{2t-1}\right)-1.
\end{align*}
Then, 
\begin{equation}
\label{eqn: TimeAvSpGOE}
    \overline{\langle S_P(t) \rangle} =\frac{4}{3}\overline{S_{P}} N{\cal{I}}_{1}(u) - \frac{1}{3}\overline{S_{P}}{\cal{I}}_{2}(v) + \overline{S_{P}},
\end{equation}
where $u=2\Gamma t$, $v=\Gamma t/2N$, 
\begin{align} 
\label{eqn: I1_TimeAvSpGOE}
\nonumber {\cal{I}}_{1}(u) &=\frac{1}{u}\int_{0}^{u}\frac{{\cal{J}}_{1}^{2}(x)}{x^{2}}dx\\
&= \frac{2\left[{\cal{J}}_{0}^{2}(u)+{\cal{J}}_{1}^{2}(u)\right]}{3}-\frac{2{\cal{J}}_{0}(u){\cal{J}}_{1}(u)}{3u}-\frac{{\cal{J}}_{1}^{2}(u)}{3u^{2}} ,
\end{align}
and
\begin{equation}
\label{eqn: I2_TimeAvSpGOE}
{\cal{I}}_{2}(v) =\frac{1}{v}\int_{0}^{v}b_{2}(x)dx = F(v)\Theta(1-v)+G(v)\Theta(v-1)  , 
\end{equation}
where
\begin{align*}
\label{eqn: I2a_TimeAvSpGOE}
F(v) &= \frac{5}{4}(1-v)+\left( \frac{v}{2}-\frac{1}{8v} \right)\text{ln}(1+2v) ,\\
G(v) &= \frac{1}{2v}(1-v)+\left( \frac{v}{2}-\frac{1}{8v} \right)\text{ln}\left(1+\frac{2}{2v-1}\right) .
\end{align*}

To obtain an expression for the asymptotic behavior of $\overline{\langle S_P(t) \rangle}$ in Eq.~(\ref{eq:SatTimeAvSp_Analytical_GOE}), we evaluate ${\cal{I}}_{1}(u\to \infty)$ and ${\cal{I}}_{2}(v\to \infty)$ given in Eqs.~(\ref{eqn: TimeAvSpGOEAsymptitic})-(\ref{eqn: TimeAvSpGOEAsymptitic02}).

\begin{figure}[h]
    \centering
\includegraphics[width=0.9\linewidth]{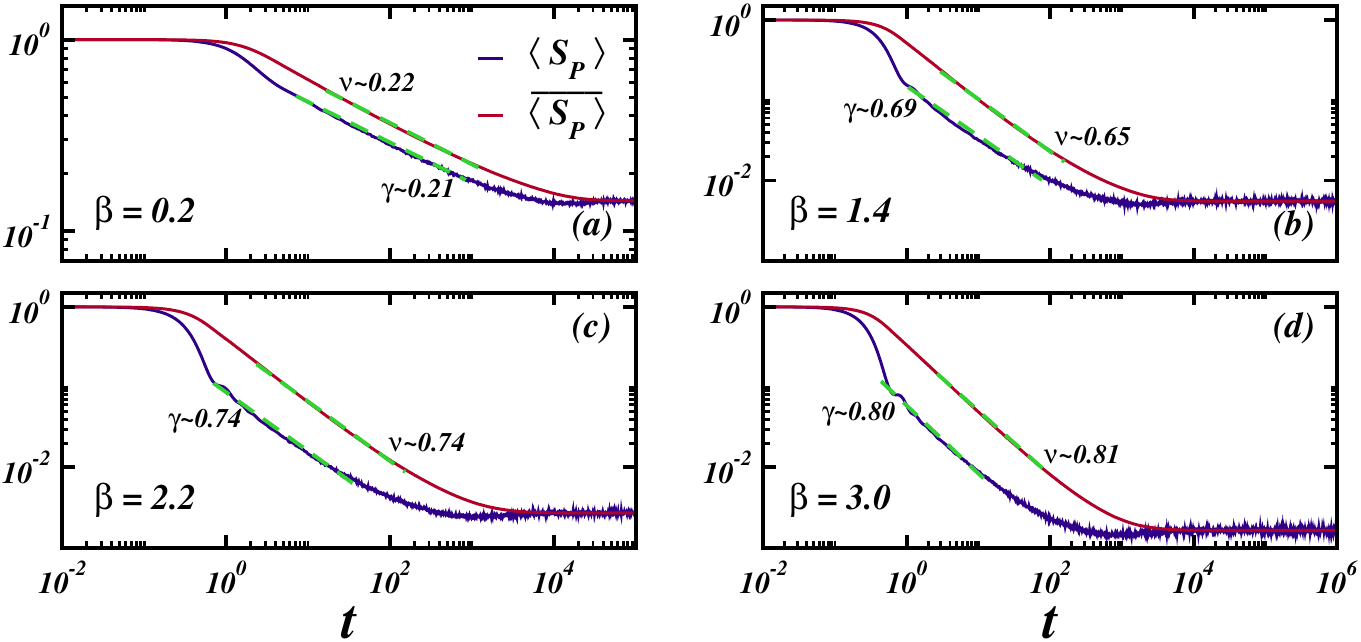}  \\
\includegraphics[width=0.6\linewidth]{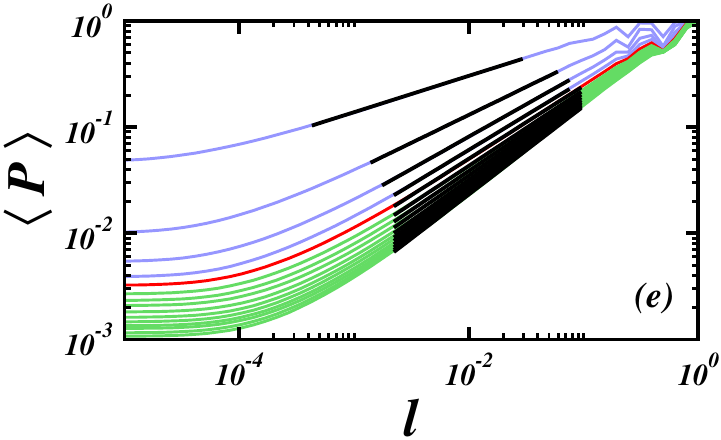}  
    \caption{Power-law banded random matrix model at the critical point $\alpha =1$ and $N=12\,000$. (a)-(d): Survival probability $\langle S_P(t)\rangle$ (dark blue) and time-averaged survival probability $\overline{\langle S_P(t)\rangle}$ (dark red). The power-law decays $\propto t^{-\gamma}$ and $\propto t^{-\nu}$ are indicated with green dashed lines and the values of $\gamma$ and $\nu$ are shown in each panel. The time intervals chosen for the fittings are the same for both exponents. The values of the bandwidth $\beta$ are also indicated.
    (e): Scaling analysis of $\langle P(l)\rangle$ from Eq.~(\ref{eq:gamma_K}) to obtain $D_2^{\text{box}}$. The blue lines are for $\beta<1$, the red curve corresponds to $\beta=1$, the green lines are for $\beta>1$, and the black lines are fitting curves.}
    \label{fig:09}
\end{figure}

\section{Additional results for power-law banded random matrices and spin-1/2 models}
{\label{Appendix B}}

Here, we provide representative figures that illustrate how (i) the power-law exponent $\gamma$ is obtained from the analysis of the decay of the survival probability, $\langle S_P(t)\rangle  \propto t^{-\gamma}$, (ii) the power-law exponent $\nu$ is obtained from the analysis of the decay of the time-averaged survival probability, $\overline{\langle S_P(t) \rangle}\propto t^{-\nu}$, and (iii) the fractal dimension $D_2^{\text{box}}$ is extracted from the scaling analysis of $\langle P (l) \rangle$. These results are shown in Fig.~\ref{fig:08} for the interacting physical spin models and in Fig.~\ref{fig:09} for the PBRM model.

The left and right panels displayed in Figs.~\ref{fig:08}(a)-(d)  show the evolution of the survival probability $\langle S_P(t)\rangle$ and of the time-averaged survival probability $\overline{\langle S_P(t)\rangle}$ for the interacting Aubry-Andr\'e model (left panels) and the disordered Heisenberg model (right panels). Both models exhibit power-law decays for both quantities at intermediate times. The bottom left and right Fig.~\ref{fig:08}(d) show the analysis of $\langle P(l) \rangle$ from Eq.~\eqref{eq:gamma_K}. The slope of the black curves for $\langle P(l) \rangle$ decreases  as the disorder strength $h$ increases, resulting in smaller values of $D_2^{\text{box}}$ as one moves away from the chaotic region.

Figure~\ref{fig:09} shows the evolution of the survival probability $\langle S_P(t)\rangle$ and of the time-averaged survival probability $\overline{\langle S_P(t)\rangle}$ for PBRM matrices  at the critical point $\alpha =1$ and for different values of the bandwidth $\beta$. Contrary to the chaotic region, where $\langle S_P(t)\rangle$ for the PBRM model shows oscillations during its power-law decay as in Fig.~\ref{fig:RMT} of the GOE model, at the critical point, $\langle S_P(t)\rangle$ decays smoothly and similarly to $\overline{\langle S_P(t)\rangle}$. Indeed, we can see in Figs.~\ref{fig:09}(a)-(d) that the decay rates for both quantities coincide, that is, $\gamma\sim\nu$. 

Figure~\ref{fig:09}(e) shows the analysis of $\langle P(l) \rangle$ from Eq.~\eqref{eq:gamma_K} for the PBRM model at the critical point $\alpha =1$ for different values of the bandwidth $\beta$. To extract the fractal dimension, we concentrate on the region indicated with black lines. The slope is larger and $D_2^{\text{box}}$ is closer to 1 for larger $\beta$.


%

\end{document}